\documentclass[12pt]{article}
\usepackage{graphicx}
\usepackage{amssymb}
\usepackage{amsmath}
\usepackage{color}
\begin{document}
\baselineskip 0.6cm
%
\begin{titlepage}
\begin{center}

\begin{flushright}
SU-HET-04-2016
\end{flushright}

\vskip 2cm

{\Large \bf 
Multiple-point principle with a scalar singlet extension
 of the Standard Model
}

\vskip 1.2cm

{\large 
Naoyuki Haba$^1$, Hiroyuki Ishida$^1$, Nobuchika Okada$^2$,\\
 and Yuya Yamaguchi$^{1,3}$
}

\vskip 0.4cm

$^1$Graduate School of Science and Engineering, Shimane University,\\
 Matsue 690-8504, Japan\\
$^2$Department of Physics and Astronomy, University of Alabama,\\
Tuscaloosa, Alabama 35487, USA\\
$^3$Department of Physics, Faculty of Science, Hokkaido University,\\
 Sapporo 060-0810, Japan

\vskip 2cm

\begin{abstract}
We suggest a scalar singlet extension of the standard model,
 in which the multiple-point principle (MPP) condition
 of a vanishing Higgs potential at the Planck scale is realized. 
Although there have been lots of attempts to realize the MPP at the Planck scale, 
 the realization with keeping naturalness is quite difficult.  
Our model can easily achieve the MPP at the Planck scale 
 without large Higgs mass corrections.
It is worth noting that the electroweak symmetry can be radiatively broken in our model.
In the naturalness point of view,
 the singlet scalar mass should be of ${\cal O}(1)\,{\rm TeV}$ or less.
We also consider right-handed neutrino extension of the model for neutrino mass generation.
The model does not affect the MPP scenario,
 and might keep the naturalness with the new particle mass scale beyond TeV,
 thanks to accidental cancellation of Higgs mass corrections.

\end{abstract}
\end{center}
\end{titlepage}

\section{Introduction}
The observed mass of the Higgs boson may imply 
 that Higgs self-coupling vanish at a high energy scale
 in the framework of the standard model (SM). 
About twenty years ago, 
 Ref.~\cite{Froggatt:1995rt} suggested the multiple-point principle (MPP) at the Planck scale, 
 and predicted Higgs boson mass as $135\pm 9\,{\rm GeV}$
 with $173\pm 5\,{\rm GeV}$ for the top quark mass.
The MPP means that there are two degenerate vacua in the SM Higgs potential,
 $V(v_H)= V(M_{\rm Pl})= 0$ and $V'(v_H) = V'(M_{\rm Pl}) = 0$,
 where $V$ is the effective Higgs potential,
 $v_H=246\,{\rm GeV}$ is the vacuum expectation value (VEV) of the Higgs doublet,
 and $M_{\rm Pl}=2.44\times 10^{18}\,{\rm GeV}$ is the reduced Planck scale. 
One is our vacuum at the electroweak (EW) scale,
 and the other vacuum lies at the Planck scale,
 which can be realized by the Planck-scale boundary conditions of
 vanishing effective Higgs self-coupling, $\lambda_H(M_{\rm Pl})=0$,
 and its beta function, $\beta_{\lambda_H}(M_{\rm Pl})=0$.
Furthermore, an asymptotic safety scenario of gravity~\cite{Shaposhnikov:2009pv}
 predicted 125\,GeV Higgs boson mass with a few GeV uncertainty.
This scenario also pointed out
 $\lambda_H(M_{\rm Pl}) \sim 0$ and $\beta_{\lambda_H}(M_{\rm Pl}) \sim 0$
 (see also Refs.~\cite{Froggatt:2001pa}-\cite{Spencer-Smith:2014woa} for more recent analyses).

Although Ref.~\cite{Froggatt:1995rt} was able to predict the approximate Higgs boson mass,
 the MPP condition can not fit the observed 125\,GeV Higgs boson mass with the recent data inputs.
In fact, within the context of the SM,
 the MPP condition at the Planck scale leads to the Higgs boson mass as $129.1\pm 1.5\,{\rm GeV}$
 by using $173.10\pm 0.59_{\rm exp} \pm 0.3_{\rm th}\,{\rm GeV}$ for the world-averaged top quark mass~\cite{Buttazzo:2013uya}.
There have been lots of attempts to realize the MPP at the Planck scale so far~\cite{Bezrukov:2014bra}-\cite{Kannike:2015apa}. 
For example, in Ref.~\cite{Haba:2014sia} the MPP at the Planck scale is achieved
 by introducing a scalar dark matter and a large Majorana mass of the right-handed neutrino. 
In this case,
 masses of the dark matter and the right-handed neutrino can be predicted.
However, there is a tension from the view point of naturalness,
 since the Higgs mass corrections via the heavy particles well-exceeds the EW scale.
Actually, it turns out to be quite difficult
 to realize the MPP at the Planck scale while keeping naturalness.

The difficulty is related with the renormalization group (RG) running of the Higgs self-coupling.
In order to satisfy $\lambda_H(M_{\rm Pl})=0$ and $\beta_{\lambda_H}(M_{\rm Pl})=0$ simultaneously,
 there should exist one or more new particles
 which change $\beta_{\lambda_H}$ adequately from the SM case.
In almost all cases,
 such new particles need to be much heavier than the EW scale,
 as long as the Higgs self-coupling is ``continuous" during the RG running.
However,
 when a new scalar field couples with the Higgs doublet and develops nonzero VEV,
 the Higgs self-coupling has a tree-level threshold correction~\cite{Gogoladze:2008gf}-\cite{Haba:2016zbu}.\footnote{
When a new heavy fermion couples with the Higgs doublet,
 there is a one-loop threshold correction,
 but it is usually negligibly small.}
The correction causes a gap between the Higgs self-coupling in the extended model and the one in the effective theory,
 which is identified as the SM one.
It has been shown that
 using the gap,
 the EW vacuum can be stabilized
 in a scalar singlet extended model~\cite{EliasMiro:2012ay}
 and type-II seesaw model~\cite{Haba:2016zbu}.
Most importantly, even if the new scalar particle is as light as a TeV scale,
 the gap can appear.
Then, the model does not affect the naturalness in the sense of Bardeen~\cite{Bardeen:1995kv}.

Here, we comment on the naturalness.
According to the Bardeen's argument,
 in quantum corrections quadratic divergences can be treated as an unphysical quantity,
 so that only logarithmic divergences should be concerned.
In this sense,
 there is no hierarchy problem within the SM,
 which possesses an approximate scale invariance
 and its stability is guaranteed by the smallness of logarithmic corrections.
Since the logarithmic corrections can be taken into account as a beta function of the Higgs mass parameter,
 the naturalness can be evaluated with the solution of its RG equation.
Namely, it is natural if the Higgs mass parameter does not significantly change during the RG running.
We will apply this sense of naturalness to our model.

In this paper,
 we will investigate the MPP condition in a scalar singlet extended model,
 which can be consistent with the 125\,GeV Higgs boson mass.
Our model is explained in the next section,
 in which we show the gap explicitly.
Numerical analyses of the MPP scenario are given in Sec.~\ref{Sec:MPP}.
We will find that the EW symmetry can be radiatively broken in our model.
We also discuss the naturalness of the Higgs mass.
In Sec.~\ref{Sec:NR},
 we will introduce right-handed neutrinos into the scalar singlet extended model
 to incorporate active neutrino masses.
We will show that in the presence of the right-handed neutrinos,
 the MPP scenario can be realized.
It will be pointed out that
 even if the singlet scalar and the right-handed neutrinos are much heavier than the EW scale,
 the model might keep the naturalness 
 thanks to an accidental cancellation of Higgs mass corrections coming from them.
Finally, we will summarize our results in Sec.~\ref{Sec:Summary}.

\section{Scalar singlet extension}
We consider a simple extension of the SM with a real singlet scalar field.
The scalar potential is given by~\cite{Egana-Ugrinovic:2015vgy}
\begin{eqnarray}
	V(H, S) = \frac{\lambda_H}{2} (H^\dagger H)^2 + m_H^2 H^\dagger H
		+ \frac{\lambda_S}{8} S^4 + \frac{\mu_S}{3} S^3 + \frac{m_S^2}{2} S^2
		+ \frac{\lambda_{HS}}{2} S^2 H^\dagger H + \mu_{HS} S H^\dagger H,
\label{potential}
\end{eqnarray}
 where $H$ and $S$ are the Higgs doublet and the scalar singlet fields, respectively.
In this paper, we consider the case with $m_S^2 > |m_H^2|$ and $\mu_{HS} > 0$,
 and omit a linear term of the singlet scalar field, which can vanish by a shift of the field.
Note that we do not assume an ad hoc $Z_2$ symmetry,
 and then, we will find that $\mu_{HS}$ plays an important role
 for the vacuum stability and the EW symmetry breaking.
In the unitary gauge, the scalar fields are written by
\begin{eqnarray}
	H = \left( 0,\ \frac{v_H+h}{\sqrt{2}} \right)^T,\quad
	S = v_S + s,
\end{eqnarray}
 where $v_H$ and $v_S$ are vacuum expectation values.
The Higgs VEV is $v_H=246\,{\rm GeV}$,
 and $v_S$ has a negative small value in our setup as will be discussed below.

The minimization conditions of the potential are given by
\begin{eqnarray}
	\frac{\partial V}{\partial h}\Bigr|_{h \to 0,\, s \to 0} &=& \frac{v_H}{2} \left(
		\lambda_H v_H^2 + 2 m_H^2 + \lambda_{HS} v_S^2 + 2 \mu_{HS} v_S \right) = 0, \label{vH}\\
	\frac{\partial V}{\partial s}\Bigr|_{h \to 0,\, s \to 0} &=& \frac{1}{2} \left[
		v_S \left( \lambda_S v_S^2 + 2 \mu_S v_S + 2 m_S^2 + \lambda_{HS} v_H^2 \right) + \mu_{HS} v_H^2 \right] = 0 . \label{vS}
\end{eqnarray}
From Eq.\,(\ref{vH}), the Higgs VEV is obtained by
\begin{eqnarray}
	v_H^2 = - \frac{1}{\lambda_H} \left( 2 m_H^2 + \lambda_{HS} v_S^2 + 2 \mu_{HS} v_S \right).
\end{eqnarray}
To realize the EW symmetry breaking,
 the Higgs mass term $m_H^2$ is negative at the EW scale,
 and $2 (- m_H^2) > \lambda_{HS} v_S^2 + 2 \mu_{HS} v_S$ should be satisfied.
Without any fine-tuning,
 we can expect $\mu_{HS} \simeq m_S$ by a naive dimensional analysis.
Thus, $|v_S|$ should be much smaller than $v_H$ for $m_S^2 \gg |m_H^2|$.

The nonzero Higgs VEV induces a tadpole for the singlet scalar due to the $\mu_{HS}$ term.
If we neglect the cubic term of $S$,
 Eq.\,(\ref{vS}) is approximated by $m_S^2 v_S + \mu_{HS} v_H^2 \approx 0$
 for $\lambda_S \leq {\cal O}(1)$, $\lambda_{HS} \leq {\cal O}(1)$.
It gives the singlet VEV as
\begin{eqnarray}
	v_S \approx - \frac{\mu_{HS} v_H^2}{2 m_S^2},
\label{vS_app}
\end{eqnarray}
 and its order of magnitude is ${\cal O}(v_H^2/m_S)$ for $\mu_{HS} \simeq m_S$.
In the no tadpole limit $\mu_{HS} \to 0$, $v_S$ vanishes.
The assumption of $\mu_S = 0$ seems to be unnatural,
 but it is necessarily required by the MPP condition as discussed later.
Actually, we will find that $\lambda_S$ and $\lambda_{HS}$ also vanish by the MPP condition.

The mass matrix for the scalar fields is expressed by the second derivatives of the potential at the VEVs:
\begin{eqnarray}
	(h,\ s) \left( \begin{array}{cc} m_{hh}^2 & m_{hs}^2 \\ m_{hs}^2 & m_{ss}^2 \end{array} \right)
	\left( \begin{array}{c} h \\ s \end{array} \right) 
	= (\phi_1,\ \phi_2) \left( \begin{array}{cc} m_{\phi_1}^2 & 0 \\ 0 & m_{\phi_2}^2 \end{array} \right)
	\left( \begin{array}{c} \phi_1 \\ \phi_2 \end{array} \right)
\end{eqnarray}
 with
\begin{eqnarray}
	\frac{\partial^2 V}{\partial h^2}\Bigr|_{h \to 0,\, s \to 0} &=& m_{hh}^2
		= \frac{3}{2} \lambda_H v_H^2 + m_H^2 + \frac{1}{2} \lambda_{HS} v_S^2 + \mu_{HS} v_S, \label{mhh}\\
	\frac{\partial^2 V}{\partial h \partial s}\Bigr|_{h \to 0,\, s \to 0} &=& m_{hs}^2
		= \lambda_{HS} v_H v_S + \mu_{HS} v_H, \label{mhs} \\
	\frac{\partial^2 V}{\partial s^2}\Bigr|_{h \to 0,\, s \to 0} &=& m_{ss}^2
		= \frac{3}{2} \lambda_S v_S^2 + 2 \mu_S v_S + m_S^2 + \frac{1}{2} \lambda_{HS} v_H^2, \label{mss}
\end{eqnarray}
and
\begin{eqnarray}
	m_{\phi_1}^2 &=& \frac{1}{2} \left( m_{hh}^2 + m_{ss}^2 - \sqrt{(m_{hh}^2 - m_{ss}^2)^2 + 4 m_{hs}^4} \right), \\
	m_{\phi_2}^2 &=& \frac{1}{2} \left( m_{hh}^2 + m_{ss}^2 + \sqrt{(m_{hh}^2 - m_{ss}^2)^2 + 4 m_{hs}^4} \right).
\end{eqnarray}
We identify the lighter eigenstate $\phi_1$ with the SM-like Higgs,
 and its mass eigenvalue $m_{\phi_1}$ corresponds to the observed Higgs boson mass $M_h = 125\,{\rm GeV}$.
In our numerical calculation,
 we will take into account a renormalization group effect for the Higgs mass.
The scalar-mixing matrix is defined by
\begin{eqnarray}
	\left( \begin{array}{c} \phi_1 \\ \phi_2 \end{array} \right) =
		\left( \begin{array}{cc} \cos \alpha & - \sin \alpha \\ \sin \alpha & \cos \alpha \end{array} \right)
		\left( \begin{array}{c} h \\ s \end{array} \right) \quad
	{\rm with} \quad
	\tan 2 \alpha = \frac{2 m_{hs}^2}{m_{ss}^2 - m_{hh}^2}.
\label{tan_a}
\end{eqnarray}
For $|m_H^2| \ll m_S^2 \simeq \mu_S^2$, 
 the mixing coupling is obtained by $\sin \alpha \approx \mu_{HS} v_H/m_S^2$,
 and it must be lower than the experimental bound $|\sin \alpha| \leq 0.36$ given by the LHC Run 1 data~\cite{LHC}.
This constraint induces $m_S \simeq \mu_{HS} \gtrsim 685\,{\rm GeV}$,
 and also $|v_S| \lesssim 45\,{\rm GeV}$ from Eq.\,(\ref{vS_app}).


In the low energy effective theory,
 the tree-level effective Higgs potential is given by~\cite{Egana-Ugrinovic:2015vgy}
\begin{eqnarray}
	V_{\rm eff} (H) = m_{\rm SM}^2 H^\dagger H + \frac{1}{2} \lambda_{\rm SM} (H^\dagger H)^2
		+ \frac{1}{3} \eta_6 (H^\dagger H)^3 + \frac{1}{8} \eta_8 (H^\dagger H)^4,
\end{eqnarray}
 with
\begin{eqnarray}
	m_{\rm SM}^2 = m_H^2, \quad
	\lambda_{\rm SM} = \lambda_H - \frac{\mu_{HS}^2}{m_S^2}, \quad
	\eta_6 = \frac{3 \lambda_{HS} \mu_{HS}^2}{2 m_S^4} - \frac{\mu_S \mu_{HS}^3}{m_S^6}, \quad
	\eta_8 = \frac{\lambda_S \mu_{HS}^4}{2 \mu_S^8}.
\end{eqnarray}
Note that the Higgs self-coupling has a nontrivial gap $\Delta \lambda \equiv \mu_{HS}^2/m_S^2$.
It can play a crucial role to make the EW vacuum stable like in a scenario in Refs.~\cite{EliasMiro:2012ay,Haba:2016zbu}.
In particular, the Higgs self-coupling $\lambda_H$ can vanish at the UV scale, e.g. the Planck scale,
 as well as the effective Higgs self-coupling $\lambda_{\rm SM}$ explains the observed Higgs boson mass,
 which has been studied in a type-II seesaw model~\cite{Gogoladze:2008gf}.
This scenario indicates
\begin{eqnarray}
	\lambda_H (M_{\rm Pl}) = 0 \quad {\rm and} \quad
	\lambda_{\rm SM} (v_H) = \frac{M_h^2}{v_H^2}.
\label{BC}
\end{eqnarray}
We show the RG running of the Higgs self-coupling in Fig.~\ref{Fig:lam_run},
 where we have used the beta functions given in Appendix~\ref{app:RGE}.
\begin{figure}[t]
\begin{center}
	\includegraphics[width=8cm,clip]{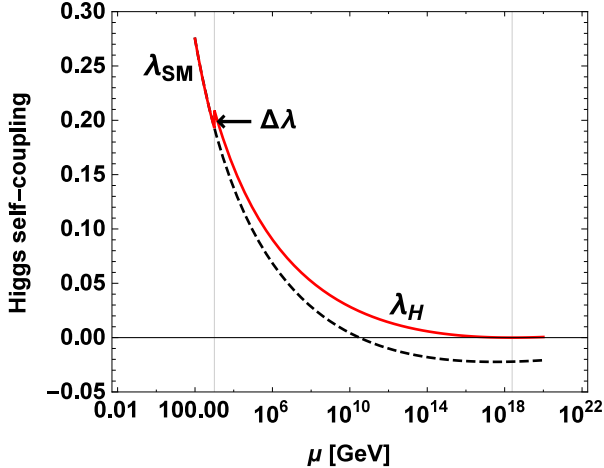}
	\caption{Renormalization group running of the Higgs self-coupling in our model (Red).
		The black-dashed line shows the running of the Higgs self-coupling in the SM.
		The vertical lines correspond to $m_S = 1\,{\rm TeV}$ and $M_{\rm Pl}$, respectively.
		We have used $M_h=125.09\,{\rm GeV}$, $M_t=172.687\,{\rm GeV}$ and $\alpha_s=0.1185$
 as the reference values.}
\label{Fig:lam_run}
\end{center}
\end{figure}
The vertical and horizontal axes show the Higgs self-coupling and renormalization scale $\mu$, respectively.
Here, we have considered $m_S$ as the cutoff of the SM,
 and taken the boundary condition $\lambda_{\rm SM} = \lambda_H - \Delta \lambda$
 at $\mu = m_S = 1\,{\rm TeV}$.
Figure~\ref{Fig:lam_run} shows that the Higgs self-coupling remains positive up to the Planck scale,
 and thus, the EW vacuum can be stabilized.

\section{Multiple-point principle} \label{Sec:MPP}

The MPP condition requires vanishing all scalar-quartic couplings and simultaneously vanishing their beta functions
 at the UV scale.
In particular, $\beta_{\lambda_H}(M_{\rm Pl})=0$ with $\lambda_H(M_{\rm Pl}) = 0$ requires
 the top Yukawa coupling as $y_t(M_{\rm Pl}) \simeq 0.388$.
In this paper, when we solve the RG equations,
 we use boundary conditions Eqs.~(\ref{BC_gY})--(\ref{BC_alpha}).
Then, to realize $y_t(M_{\rm Pl}) \simeq 0.388$,
 the top pole mass $M_t$ should be taken as
 172.322\,GeV, 172.687\,GeV and 173.052\,GeV
 for the fixed strong coupling $\alpha_s(M_Z)=0.1179$, 0.1185 and 0.1191, respectively.
For measurements of the top pole mass, 
 $M_t = 172.99 \pm 0.91\,{\rm GeV}$~\cite{Aad:2015nba} and
 $M_t = 172.44 \pm 0.48\,{\rm GeV}$~\cite{Khachatryan:2015hba}
 are obtained by the ATLAS and CMS collaborations, respectively.
Thus, our result expected by the MPP is consistent with the current experimental data.
In the following, we take $\alpha_s(M_Z)=0.1185$ and $M_t=172.687\,{\rm GeV}$ as reference values.

Imposing the MPP condition in the scalar singlet extended model,
 $\lambda_S$ and $\lambda_{HS}$ remain zero during the RG runnings.
Then, the MPP condition also requires
 a vanishing triple coupling of the singlet scalar ($\mu_S$),
 because the highest term of $S$ must be even function to realize the degenerate vacua.
Once $\mu_S$ vanishes, it also remains zero.
In the rest of this paper,
 we can take away $\lambda_S$, $\lambda_{HS}$ and $\mu_S$ from our discussion.
Note that for the vacuum around the Planck scale, $v_H \sim M_{\rm Pl}$,
 the stationary condition~(\ref{vS}) suggests
 $v_S \sim M_{\rm Pl}$ with $\lambda_S(M_{\rm Pl}) \sim \mu_{HS}(M_{\rm Pl}) / M_{\rm Pl}$.
This value of $\lambda_S(M_{\rm Pl})$ is extremely small
 and practically we can use the MPP condition as $\lambda_S (M_{\rm Pl})=0$.

It is worth noting that $\Delta \lambda$ is uniquely determined for a given $m_S$,
 once the MPP condition and Eq.~(\ref{BC}) are required.
Then, $\mu_{HS}$ is determined by $\mu_{HS}^2 = \Delta \lambda \, m_S^2$.
In addition, $v_S$ is exactly obtained by Eq.~(\ref{vS_app})
 because of $\lambda_S=\lambda_{HS}=0$ and $\mu_S=0$.
As a result, our model is controlled by only one free parameter.
In the following,
 we choose $m_S$ as the free parameter.

\begin{figure}[t]
\begin{center}
	\includegraphics[width=8cm,clip]{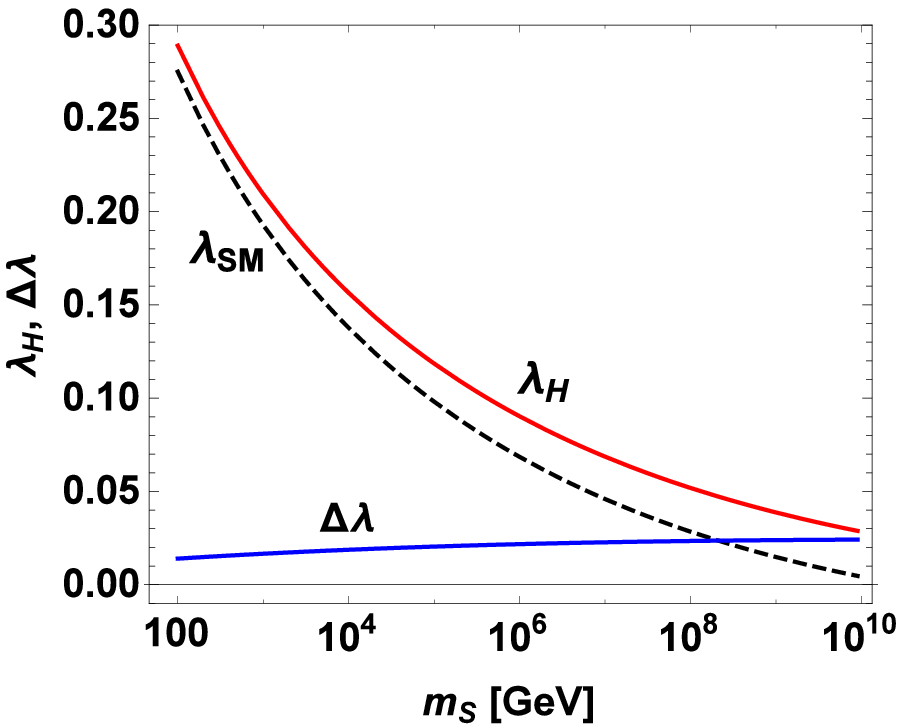} \hspace{5mm}
	\includegraphics[width=8cm,clip]{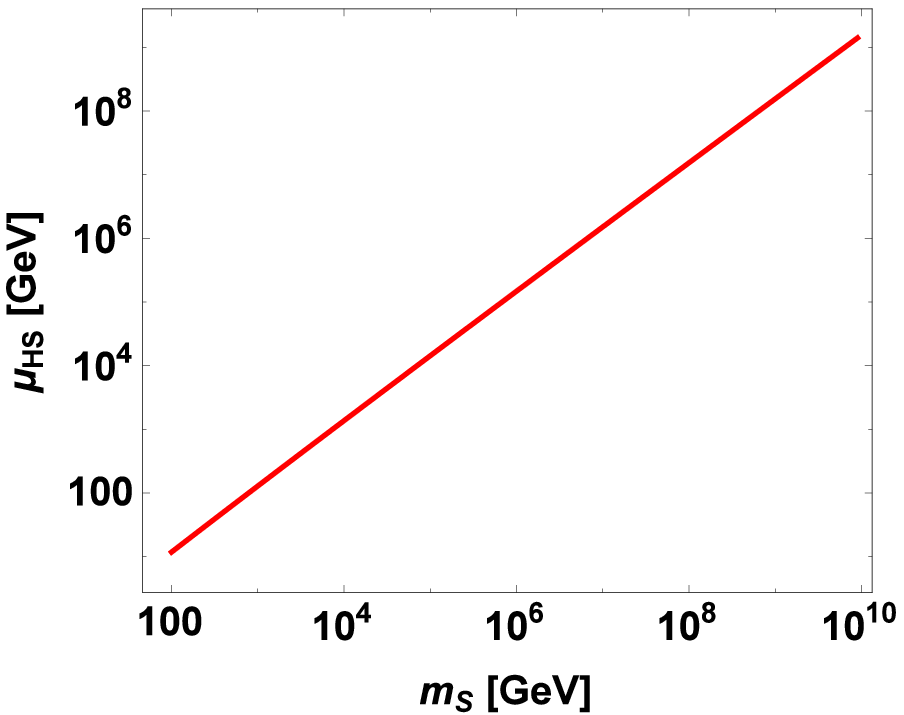}
	\caption{Left: $m_S$ dependence of $\Delta \lambda$ (Blue).
		The red and black-dashed line show $\lambda_H(m_S)$ and $\lambda_{\rm SM}(m_S)$, respectively.
		Right: $m_S$ dependence of $\mu_{HS}$.}
\label{Fig:lambda}
\end{center}
\end{figure}

The left panel of Fig.~\ref{Fig:lambda} shows $m_S$ dependence of $\Delta \lambda$ as the blue line.
The red and black-dashed line show $\lambda_H(m_S)$ and $\lambda_{\rm SM}(m_S)$, respectively.
We find that $\Delta \lambda$ is almost constant,
 and thus, $\mu_{HS}(m_S)$ is roughly proportional to $m_S$
 as shown in the right panel of Fig.~\ref{Fig:lambda}.
To stabilize the EW vacuum,
 the Higgs self-coupling should remain positive up to the Planck scale.
Thus, $m_S$ has to be smaller than $10^{10}\,{\rm GeV}$,
 and we do not consider the heavier case.

\begin{figure}[t!]
\begin{center}
	\includegraphics[width=8cm,clip]{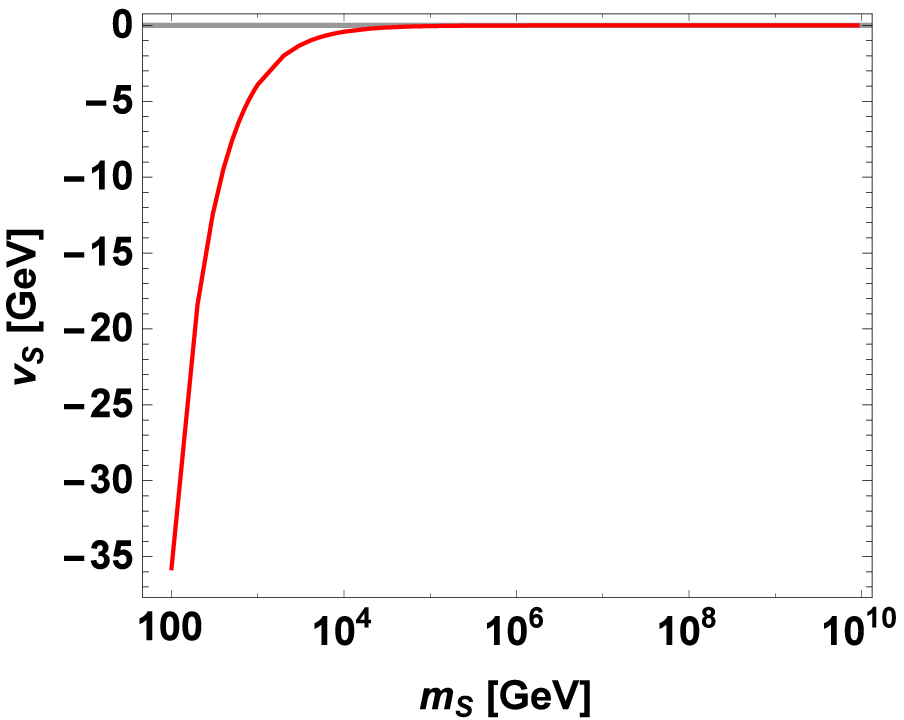} \hspace{5mm}
	\includegraphics[width=8cm,clip]{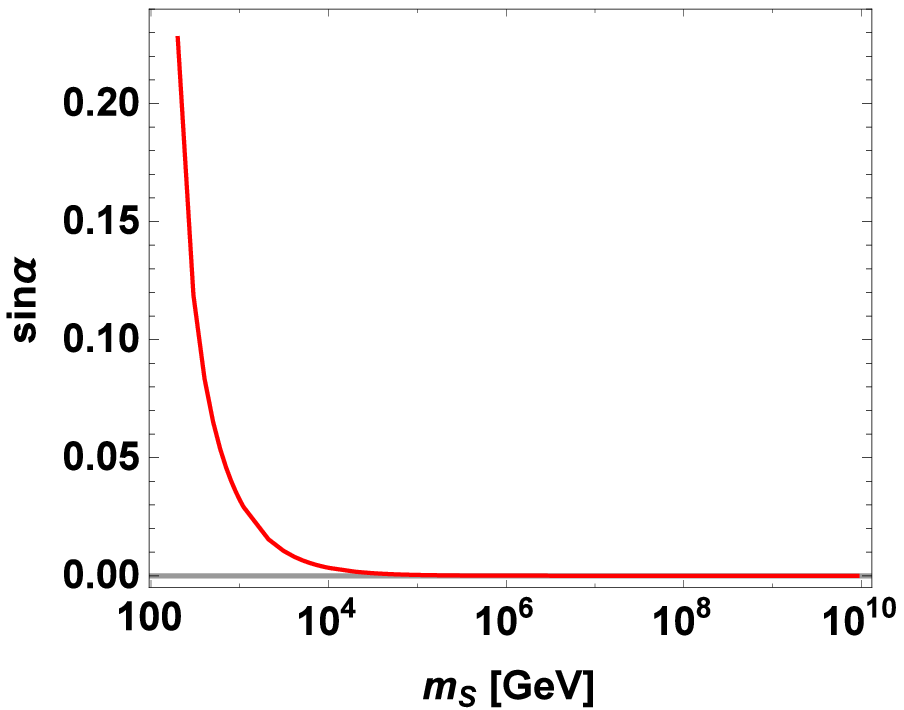}
	\caption{$v_S$ (Left) and $\sin \alpha$ (Right) as a function of $m_S$.}
\label{Fig:vS}
\end{center}
\end{figure}

Figure~\ref{Fig:vS} shows $m_S$ dependences of $v_S$ and $\sin \alpha$ in the left and right panels, respectively.
Imposing the MPP condition,
 Eq.~(\ref{vS_app}) becomes exact equal,
 where values of $v_S$ is obtained as $v_S = - \sqrt{\Delta \lambda} v_H^2/ (2 m_S)$.
Since $\Delta \lambda$ is almost constant,
 $v_S$ is almost inversely proportional to $m_S$.
We find that $-35\,{\rm GeV} \lesssim v_S < 0\,{\rm GeV}$
 and particularly $|v_S| < 1\,{\rm GeV}$ for $m_S > 4\,{\rm TeV}$.
The scalar-mixing angle is obtained by
\begin{eqnarray}
	\tan 2\alpha = \frac{2 \mu_{HS} v_H}{m_S^2 - \lambda_H v_H^2} \quad \longrightarrow \quad
	\sin \alpha \approx \alpha \approx \frac{\mu_{HS} v_H}{m_s^2} = 2 \frac{-v_S}{v_H} \quad
	{\rm for}\ m_S^2 \gg |m_H^2|.
\end{eqnarray}
Thus, we can estimate $\sin \alpha < 0.01$ for $m_S > 4\,{\rm TeV}$.
Note that all parameter region is safe from the LHC Run 1 constraint $|\sin \alpha| \leq 0.36$~\cite{LHC}.
This result is different from the estimation discussed below Eq.~(\ref{tan_a}).
The reason is that the estimation comes from $\mu_{HS} \simeq m_S$,
 while the MPP condition requires $\mu_{HS} \simeq 0.1\, m_S$.

\begin{figure}[t]
\begin{center}
	\includegraphics[width=8cm,clip]{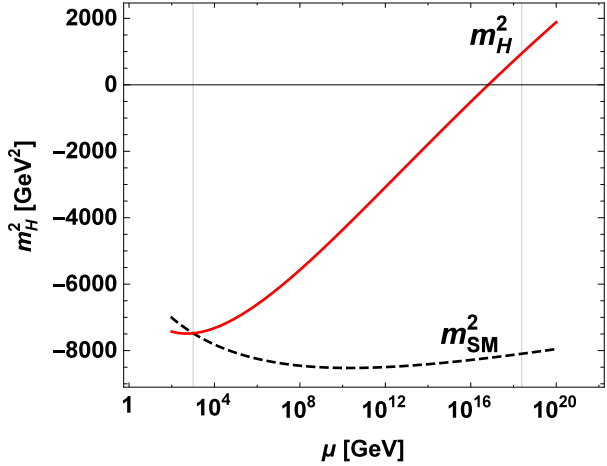}
	\caption{Renormalization group running of $m_H^2$ (Red).
		The black-dashed line shows the running of the Higgs mass parameter in the SM.
		The vertical lines correspond to $m_S = 1\,{\rm TeV}$ and $M_{\rm Pl}$, respectively.}
\label{Fig:mH}
\end{center}
\end{figure}

It is remarkable that the EW symmetry is radiatively broken in our model.
The beta function of $m_H^2$ is dominated by $\mu_{HS}^2$ term for $|m_H^2| \ll \mu_{HS}^2$.
Its RG solution is approximately given by
\begin{eqnarray}
	m_H^2 (\mu) \approx m_H^2 (M_{\rm Pl}) - \frac{\mu_{HS}^2}{16 \pi^2} \ln \left( \frac{M_{\rm Pl}}{\mu} \right)^2 \quad
	{\rm for}\ m_S \leq \mu \leq M_{\rm Pl}, 
\label{mH_run}
\end{eqnarray}
To realize the EW symmetry breaking, $m_H^2$ should be negative at the EW scale,
 while $m_H^2$ is positive at the Planck scale as $m_H^2 (M_{\rm Pl}) \sim \mu_{HS}^2$.
This behavior is explicitly shown in Fig.~\ref{Fig:mH}.
Here, we have taken the cutoff of the SM at $\mu = m_S (m_S)=1\,{\rm TeV}$,
 and then, $\Delta \lambda \simeq 0.0166$, $\mu_{HS}(m_S) \simeq 129\,{\rm GeV}$ and $v_S \simeq -3.90\,{\rm GeV}$.

\begin{figure}[t!]
\begin{center}
	\includegraphics[width=8cm,clip]{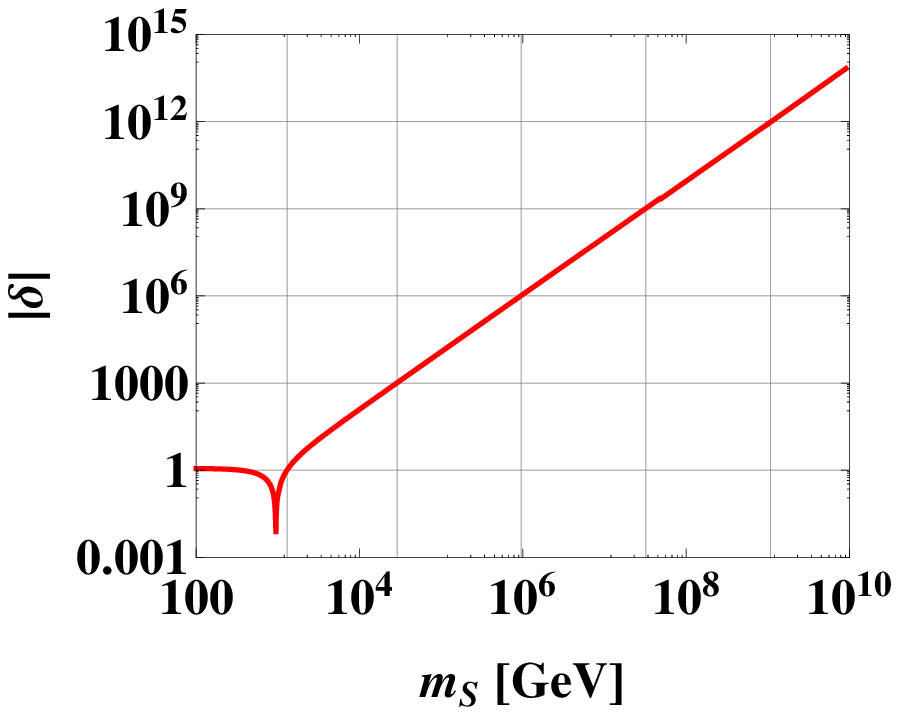} \hspace{5mm}
	\includegraphics[width=8cm,clip]{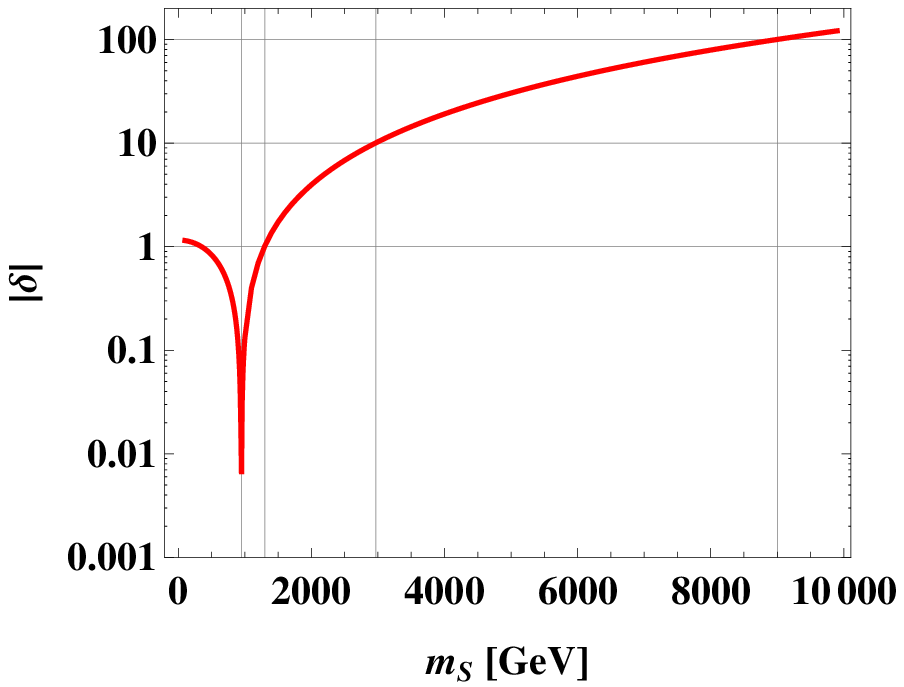}
	\caption{$m_S$ dependence of $\delta$.
		The right panel concentrates on $100\,{\rm GeV} \leq m_S \leq 10\,{\rm TeV}$.}
\label{Fig:delta_mH}
\end{center}
\end{figure}

In the end of this section,
 we mention the naturalness of the Higgs mass.
When $m_S$ is much higher than the EW scale,
 it induces $|m_H^2 (v_H)| \ll m_H^2 (M_{\rm Pl})$,
 that is, the RG running of $m_H^2$ is highly tuned to realize the observed Higgs mass.
Here, we define the fine-tuning level as
 $\delta \equiv m_H^2 (M_{\rm Pl})/|m_H^2 (v_H)|=2m_H^2 (M_{\rm Pl})/M_h^2$,
 where $M_h=125\,{\rm GeV}$.
For example, $\delta = 10$ indicates that
 we need to fine-tune the Higgs mass squared at the accuracy of 10\% level.
Figure~\ref{Fig:delta_mH} shows the $m_S$ dependence of $\delta$,
 and we find $\delta = 1$, 10 and 100 correspond to
 $m_S \simeq 1.3\,{\rm TeV}$, 3.0\,TeV and 9.0\,TeV, respectively.
Therefore, from the naturalness point of view,
 there should exist the singlet scalar at ${\cal O}(1)\,{\rm TeV}$ scale.
We have have found that $m_H^2 (M_{\rm Pl})$ vanishes for $m_S \simeq 950\,{\rm GeV}$,
 and becomes negative in the lower $m_S$ region,
 in which the radiative EW symmetry breaking does not occur.
For a tadpole diagram which contributes Higgs mass correction,
 it is tiny due to the heavy mass of $m_s$.\footnote{
For $\mu_S \neq 0$, there is a finite Higgs mass correction by a tadpole diagram of the singlet scalar.
However, we need not consider it because of $\mu_S = 0$ coming from the MPP condition.}

\section{Additional extension with right-handed neutrinos} \label{Sec:NR}

In addition to the singlet scalar,
 we can introduce right-handed neutrinos to explain the active neutrino masses.
Interaction parts of the Lagrangian including right-handed neutrinos are given by
\begin{eqnarray}
	-{\cal L}_N = Y_\nu^\dagger \overline{L} \tilde{H} N + Y_N S \overline{N} N + \frac{1}{2} M_N \overline{N^c} N + {\rm h.c.}\, , 
\end{eqnarray}
 where $L$ and $N$ are lepton doublet and right-handed neutrino fields, respectively.
Imposing the MPP condition at the Planck scale,
 $\beta_{\lambda_S}(M_{\rm Pl}) = 0$ is required,
 and then, $Y_N$ vanishes in all energy scales (see Appendix~\ref{app:RGE2}).
Therefore, new parameters are only $Y_\nu$ and $M_N$
 as same as the usual type-I seesaw model~\cite{seesaw}.
These parameters should satisfy the seesaw relation $m_\nu = Y_\nu^T M_N^{-1} Y_\nu v_H^2/2$,
 where $m_\nu$ is the active neutrino mass matrix
 calculated by mass eigenvalues and the PMNS matrix~\cite{PMNS}.

When we consider the $Y_\nu \ll {\cal O}(1)$ (or equivalently $M_N \ll {\cal O}(10^{14})\,{\rm GeV}$) case,
 right-handed neutrino contributions are negligible in runnings of the scalar-quartic couplings.
Thus, the MPP scenario remains the same as the one without right-handed neutrinos.\footnote{
When the neutrinos are Dirac fermions,
 there are no Majorana masses and $Y_\nu \ll {\cal O}(1)$.
Then, the MPP scenario can be realized as in the previous section.
}
However, only the RG running of $m_H^2$ might change significantly.
Including contributions of the right-handed neutrinos,
 Eq.~(\ref{mH_run}) is rewritten by
\begin{eqnarray}
	m_H^2 (\mu) \approx m_H^2 (M_{\rm Pl}) 
		- \frac{\mu_{HS}^2}{16 \pi^2} \ln \left( \frac{M_{\rm Pl}}{\mu} \right)^2
		+ \frac{4 N_\nu m_{\rm eff} M_N^3}{16 \pi^2 v_H^2} \ln \left( \frac{M_{\rm Pl}}{M_N} \right)^2 \quad
	{\rm for}\ m_S \leq \mu \leq M_N, 
\label{mH_run_2a}
\end{eqnarray}
or
\begin{eqnarray}
	m_H^2 (\mu) \approx m_H^2 (M_{\rm Pl}) 
		- \frac{\mu_{HS}^2}{16 \pi^2} \ln \left( \frac{M_{\rm Pl}}{m_S} \right)^2
		+ \frac{4 N_\nu m_{\rm eff} M_N^3}{16 \pi^2 v_H^2} \ln \left( \frac{M_{\rm Pl}}{\mu} \right)^2 \quad
	{\rm for}\ M_N \leq \mu \leq m_S, 
\label{mH_run_2b}
\end{eqnarray}
 where, using the seesaw relation, we have defined
 ${\rm Tr}(Y_\nu^\dagger M_N^2 Y_\nu) \equiv 2 N_\nu m_{\rm eff} M_N^3/v_H^2$
 (in the right side $M_N$ is a number not matrix).
The effective neutrino mass $m_{\rm eff}$ is typically given by the heaviest active neutrino mass,
 and $N_\nu$ means the relevant number of right-handed neutrinos.
Since the singlet scalar and the right-handed neutrinos oppositely contribute to $m_H^2$,
 the Higgs mass corrections might be accidentally canceled (at the one-loop level).

\begin{figure}[t!]
\begin{center}
	\includegraphics[width=8cm,clip]{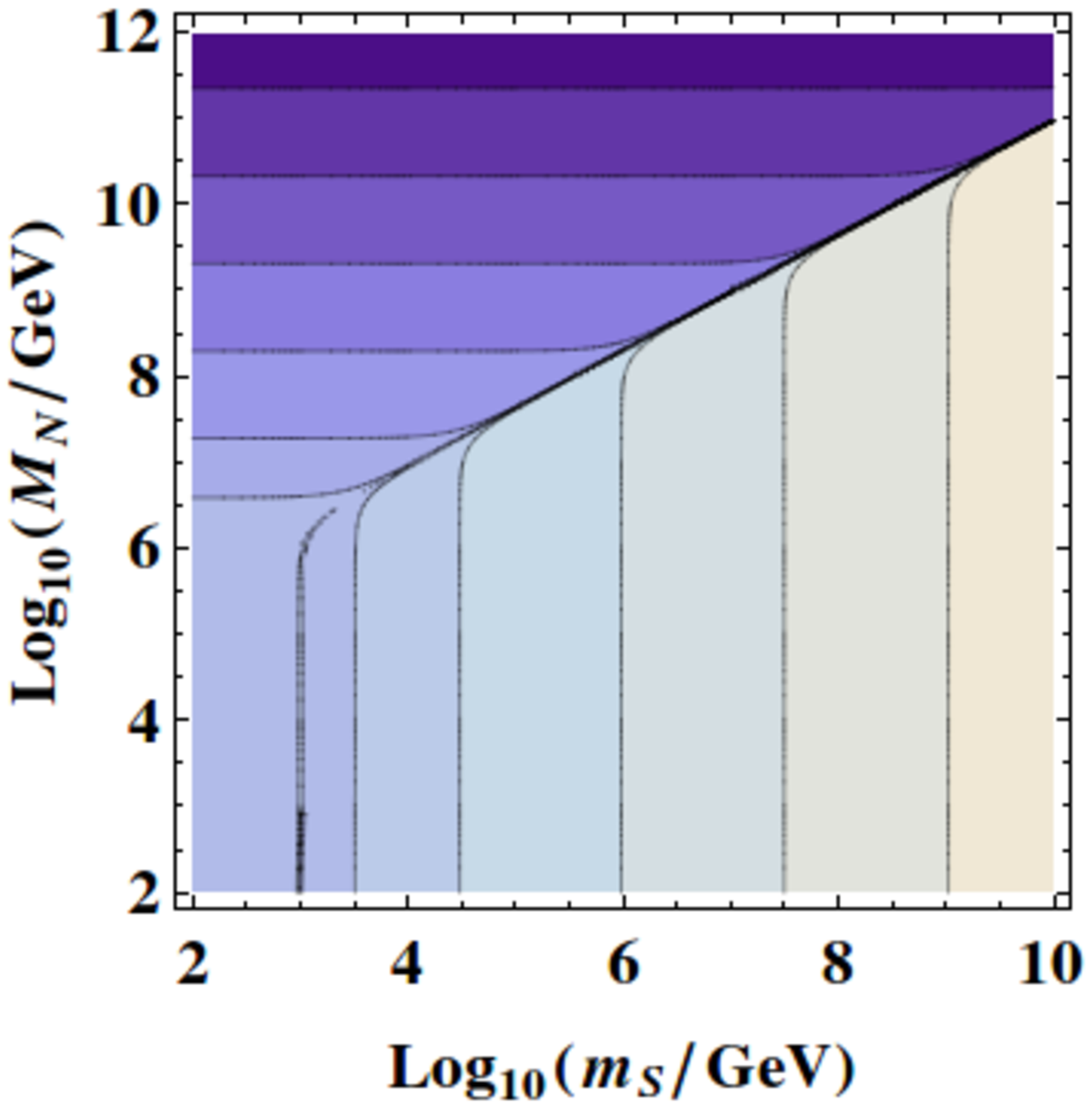} \hspace{5mm}
	\includegraphics[scale=1.4,clip]{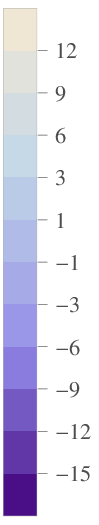}
	\caption{Contour plot of $\delta$ in ($m_S$, $M_N$) plane.
		The values of right bar are shown by ${\rm sign}[\delta]\, {\rm Log}_{10}|\delta|$,
		where we have defined ${\rm sign}[\delta] \equiv \delta/|\delta|$.}
\label{Fig:contour}
\end{center}
\end{figure}

We show contour plot of $\delta$ in Fig.~\ref{Fig:contour},
 where the horizontal and vertical axes show $m_S$ and $M_N$, respectively.
For the calculation of Eqs.~(\ref{mH_run_2a}) and (\ref{mH_run_2b}),
 we have taken $N_\nu=1$ and $m_{\rm eff} = 0.05\,{\rm eV}$ as reference values.
The positive $\delta$ region, in which singlet scalar contribution is dominant,
 can drive the radiative EW symmetry breaking
 as mentioned above.
When the right-handed neutrino mass becomes larger,
 the value of $\delta$ becomes smaller
 and vanishes at a specific point.
From Eqs.~(\ref{mH_run_2a}) and (\ref{mH_run_2b}),
 the point is estimated by
\begin{eqnarray}
	{\rm Log}_{10}\left(\frac{M_N}{{\rm GeV}}\right) \approx 4 + \frac{2}{3}{\rm Log}_{10}\left(\frac{m_S}{{\rm GeV}}\right).
\label{cancel}
\end{eqnarray}
If this relation is realized,
 $\delta$ can be small and hence our scenario can be natural
 even for the masses of singlet scalar and right-handed neutrinos $\gg 1\,{\rm TeV}$.

\section{Summary} \label{Sec:Summary}

We have investigated the scalar singlet extension of the SM with the MPP condition, 
 in which the scalar potential has two degenerate vacua at the EW and a UV scales.
The condition requires all vanishing scalar-quartic couplings and simultaneously vanishing their beta functions
 at the UV scale, which we have taken as the Planck scale.
Particularly, $\beta_{\lambda_H}(M_{\rm Pl})=0$ with $\lambda_H(M_{\rm Pl})=0$
 can determine the top pole mass as 172.322\,GeV, 172.687\,GeV and 173.052\,GeV
 for $\alpha_s(M_Z)=0.1179$, 0.1185 and 0.1191, respectively.
These values are consistent with the current experimental data
 $M_t = 172.99 \pm 0.91\,{\rm GeV}$ by the ATLAS collaboration~\cite{Aad:2015nba} and
 $M_t = 172.44 \pm 0.48\,{\rm GeV}$ by the CMS collaboration~\cite{Khachatryan:2015hba}.
The MPP conditions strongly restrict our model parameters,
 and there is only one free parameter left in our analysis,
 which we have taken the singlet mass $m_S$.
We have shown $m_S$ dependence of some model predictions,
 and found that our model is consistent with the LHC Run 1 results for the SM Higgs boson properties.

To simultaneously realize the MPP condition and the observed Higgs mass,
 singlet-Higgs-Higgs coupling $\mu_{HS}$ plays an important role.
Furthermore, this coupling induces the radiative EW symmetry breaking.
When the singlet mass is much larger than the Higgs mass,
 $\mu_{HS}^2$ term dominate the beta function of the Higgs mass squared $\beta_{m_H^2}$.
Then, the sign of $m_H^2$ can flip during the RG running,
 that is, $m_H^2$ becomes negative toward the EW scale while positive at the Planck scale.
We have found that this behavior can occur for $m_S > 950\,{\rm GeV}$.
On the other hand,
 too large $m_S$ causes the fine-tuning problem of the Higgs mass.
To avoid the problem,
 there should exist the singlet scalar at ${\cal O}(1)\,{\rm TeV}$ scale.

In order to incorporate the neutrino masses and flavor mixings to the singlet scalar extended model,
 we have introduced right-handed neutrinos and investigated the MPP scenario.
Here, new parameters $Y_\nu$ and $M_N$ are introduced,
 which are neutrino Dirac Yukawa coupling and right-handed neutrino Majorana mass matrices, respectively,
 and leading to the type-I seesaw mechanism.
For $Y_\nu \ll {\cal O}(1)$ (or equivalently $M_N \ll {\cal O}(10^{14})\,{\rm GeV}$),
 the running of all couplings except $m_H^2$ are almost the same as before.
Therefore, the model can realize the MPP scenario
 as well as explaining the active neutrino masses.

It might be possible to solve the fine-tuning problem of the Higgs mass
 by an accidental cancellation of Higgs mass corrections coming from the singlet scalar and the right-handed neutrinos.
We have found its approximate condition as Eq.~(\ref{cancel}).
If the condition is satisfied,
 masses of singlet scalar and right-handed neutrinos can exceed ${\cal O}(1)\,{\rm TeV}$.

\subsection*{Acknowledgment}
This work is partially supported by Scientific Grants
  by the Ministry of Education, Culture, Sports, Science and Technology (Nos. 24540272, 26247038, and 15H01037)
  and the United States Department of Energy (DE-SC 0013680).
The work of Y. Y. is supported
  by Research Fellowships of the Japan Society for the Promotion of Science for Young Scientists
  (Grants No. 26$\cdot$2428).

\section*{Appendix}
\appendix
\section{Beta functions in the scalar singlet extended model} \label{app:RGE}

The one-loop beta functions for the SM are given by
\begin{eqnarray}
	\beta_{g_Y} &=& \frac{g_Y^3}{16 \pi^2} \frac{41}{6}, \qquad
	\beta_{g_2} = \frac{g_2^3}{16 \pi^2} \left( -\frac{19}{6} \right), \qquad
	\beta_{g_3} = \frac{g_3^3}{16 \pi^2} \left( -7 \right), \\
	\beta_{y_t} &=& \frac{y_t}{16 \pi^2} \left( 
		- \frac{9}{4} g_2^2 - 8 g_3^2 - \frac{17}{12} g_Y^2 + \frac{9}{2} y_t^2 \right), \\
	\beta_{\lambda_{\rm SM}} &=& \frac{1}{16 \pi^2} \left[
		\lambda_{\rm SM} \left( 12 \lambda_{\rm SM} - 9 g_2^2 - 3 g_Y^2 + 12 y_t^2 \right)
		+ \frac{9}{4} g_2^4 + \frac{3}{2} g_2^2 g_Y^2 + \frac{3}{4} g_Y^4 - 12 y_t^4 \right], \\
	\beta_{m_{\rm SM}^2} &=& \frac{m_{\rm SM}^2}{16 \pi^2} \left(
		6 \lambda_{\rm SM} - \frac{9}{2} g_2^2 - \frac{3}{2} g_Y^2 + 6 y_t^2 \right).
\end{eqnarray}
Here, we omit the Yukawa couplings except for the top Yukawa coupling,
 since the other Yukawa couplings enough small to be neglected.

For a real singlet scalar extension of the SM,
 the one-loop beta functions of the gauge and the top Yukawa couplings do not change.
The beta functions of the other couplings are given by
\begin{eqnarray}
	\beta_{\lambda_H} &=& \frac{1}{16 \pi^2} \left[
		\lambda_H \left( 12 \lambda_H - 9 g_2^2 - 3 g_Y^2 + 12 y_t^2 \right)
		+ \frac{9}{4} g_2^4 + \frac{3}{2} g_2^2 g_Y^2 + \frac{3}{4} g_Y^4 - 12 y_t^4 + \lambda_{HS}^2 \right], \\
	\beta_{\lambda_S} &=& \frac{1}{16 \pi^2}  \left(
		9 \lambda_S^2 + 4 \lambda_{HS}^2 \right), \\
	\beta_{\lambda_{HS}} &=& \frac{\lambda_{HS}}{16 \pi^2} 
		\left( 6 \lambda_H - \frac{9}{2} g_2^2 - \frac{3}{2} g_Y^2 + 6 y_t^2
		+ 4 \lambda_{HS} + 3 \lambda_S \right), \\
	\beta_{\mu_S} &=& \frac{1}{16 \pi^2} \left(
		9 \lambda_S \mu_S + 6 \lambda_{HS} \mu_{HS} \right), \\
	\beta_{\mu_{HS}} &=& \frac{1}{16 \pi^2} \left[
		\mu_{HS} \left( 6 \lambda_H - \frac{9}{2} g_2^2 - \frac{3}{2} g_Y^2 + 6 y_t^2 + 4 \lambda_{HS} \right)
		+ \lambda_{HS} \mu_S \right], \\
	\beta_{m_H^2} &=& \frac{1}{16 \pi^2} \left[ m_H^2 \left(
		6 \lambda_H - \frac{9}{2} g_2^2 - \frac{3}{2} g_Y^2 + 6 y_t^2 \right)
		+ \lambda_{HS} m_S^2 + 2 \mu_{HS}^2 \right], \\
	\beta_{m_S^2} &=& \frac{1}{16 \pi^2} \left(
		3 \lambda_S m_S^2 + 4 \mu_S^2 + 4 \lambda_{HS} m_H^2 + 4 \mu_{HS}^2 \right).
\end{eqnarray}

To solve the RG equations, we take the following boundary conditions \cite{Buttazzo:2013uya,Agashe:2014kda}:
\begin{eqnarray}
	&&g_Y(M_t) = 0.35761 + 0.00011 \left( \frac{M_t}{{\rm GeV}} - 173.10 \right),
\label{BC_gY} \\ 
	&&g_2(M_t) = 0.64822 + 0.00004 \left( \frac{M_t}{{\rm GeV}} - 173.10 \right),\\
	&&g_3(M_t) = 1.1666 - 0.00046 \left( \frac{M_t}{{\rm GeV}} - 173.10 \right) + 0.00314 \left( \frac{\alpha_3(M_Z) - 0.1184}{0.0007} \right),\\
	&&y_t(M_t) = 0.93558 + 0.00550 \left( \frac{M_t}{{\rm GeV}} - 173.10 \right) -0.00042 \left( \frac{\alpha_3(M_Z) - 0.1184}{0.0007} \right),\\
	&&\alpha_s(M_Z) = 0.1185 \pm 0.0006,
\label{BC_alpha}
\end{eqnarray}
 where $M_t$ is the pole mass of top quark.
In our analysis, the top pole mass is determined by the MPP condition:
 $M_t = 172.322\,{\rm GeV}$, 172.687\,GeV and 173.052\,GeV
 for $\alpha_s(M_Z)=0.1179$, 0.1185 and 0.1191, respectively.

\newpage
\section{Beta functions in the scalar singlet extended model with right-handed neutrinos} \label{app:RGE2}

In addition to the real singlet scalar field,
 we introduce right-handed neutrinos.
The one-loop beta functions of the gauge couplings do not change.
The beta functions of the other couplings are given by
 \begin{eqnarray}
	\beta_{y_t} &=& \frac{y_t}{16 \pi^2} \left( 
		- \frac{9}{4} g_2^2 - 8 g_3^2 - \frac{17}{12} g_Y^2 + \frac{9}{2} y_t^2
		+ {\rm Tr}(Y_\nu^\dagger Y_\nu) \right), \\
	\beta_{Y_\nu} &=& \frac{1}{16 \pi^2} \left[ Y_\nu \left( 
		- \frac{9}{4} g_2^2 - \frac{3}{4} g_Y^2 + 3 y_t^2 + {\rm Tr}(Y_\nu^\dagger Y_\nu)
		+ \frac{3}{2} Y_\nu^\dagger Y_\nu  \right) + 2 Y_N^2 Y_\nu \right], \\
	\beta_{Y_N} &=& \frac{1}{16 \pi^2} \left[ Y_N \left( 
		4 {\rm Tr}(Y_N^2) +12 Y_N^2 + (Y_\nu Y_\nu^\dagger)^T  \right)
		+ Y_\nu Y_\nu^\dagger Y_N \right], \\
	\beta_{\lambda_H} &=& \frac{1}{16 \pi^2} \left[
		\lambda_H \left( 12 \lambda_H - 9 g_2^2 - 3 g_Y^2 + 12 y_t^2 + 4 {\rm Tr}(Y_\nu^\dagger Y_\nu) \right)
		+ \frac{9}{4} g_2^4 + \frac{3}{2} g_2^2 g_Y^2 + \frac{3}{4} g_Y^4 \right. \nonumber \\
		&& - 12 y_t^4 + \lambda_{HS}^2
		- 4 {\rm Tr}(Y_\nu^\dagger Y_\nu Y_\nu^\dagger Y_\nu) \biggr], \\
	\beta_{\lambda_S} &=& \frac{1}{16 \pi^2}  \left[
		\lambda_S \left( 9 \lambda_S + 16 {\rm Tr}(Y_N^2) \right) + 4 \lambda_{HS}^2
		- 128 {\rm Tr}(Y_N^4) \right], \\
	\beta_{\lambda_{HS}} &=& \frac{1}{16 \pi^2} \left[
		\lambda_{HS} \left( 6 \lambda_H - \frac{9}{2} g_2^2 - \frac{3}{2} g_Y^2 + 6 y_t^2
		+ 4 \lambda_{HS} + 3 \lambda_S
		+ 2 {\rm Tr}(Y_\nu^\dagger Y_\nu) + 8 {\rm Tr}(Y_N^2) \right) \right. \nonumber \\
		&& - 32 {\rm Tr}(Y_N^2 Y_\nu^\dagger Y_\nu) \biggr], \\
	\beta_{\mu_S} &=& \frac{1}{16 \pi^2} \left[ \mu_S \left(
		9 \lambda_S +12 {\rm Tr}(Y_N^2) \right) + 6 \lambda_{HS} \mu_{HS}
		- 96 {\rm Tr}(M_N Y_N^3) \right], \\
	\beta_{\mu_{HS}} &=& \frac{1}{16 \pi^2} \left[
		\mu_{HS} \left( 6 \lambda_H - \frac{9}{2} g_2^2 - \frac{3}{2} g_Y^2 + 6 y_t^2 + 4 \lambda_{HS}
		+ 2 {\rm Tr}(Y_\nu^\dagger Y_\nu) + 4 {\rm Tr}(Y_N^2) \right) \right. \nonumber \\
		&& + \lambda_{HS} \mu_S - 16 {\rm Tr}(M_N Y_N Y_\nu^\dagger Y_\nu) \biggr], \\
	\beta_{M_N} &=& \frac{1}{16 \pi^2} \left[
		M_N (Y_\nu Y_\nu^\dagger)^T + \left( Y_\nu Y_\nu^\dagger \right) M_N
		+ 4 {\rm Tr}(M_N Y_N)  Y_N + 12 M_N Y_N^2 \right], \\
	\beta_{m_H^2} &=& \frac{1}{16 \pi^2} \left[ m_H^2 \left(
		6 \lambda_H - \frac{9}{2} g_2^2 - \frac{3}{2} g_Y^2 + 6 y_t^2 +2 {\rm Tr}(Y_\nu^\dagger Y_\nu) \right)
		+ \lambda_{HS} m_S^2 + 2 \mu_{HS}^2 \right. \nonumber \\
		&& - 4 {\rm Tr}(Y_\nu^\dagger M_N^2 Y_\nu) \biggr], \\
	\beta_{m_S^2} &=& \frac{1}{16 \pi^2} \left[
		m_S^2 \left( 3 \lambda_S + 8 {\rm Tr}(Y_N^2) \right)
		+ 4 \mu_S^2 + 4 \lambda_{HS} m_H^2 + 4 \mu_{HS}^2
		- 48 {\rm Tr}(M_N^2 Y_N^2) \right],
\end{eqnarray}
where $Y_N$ and $M_N$ are real diagonal matrices.
We have used SARAH~\cite{Staub:2008uz} to obtain these beta functions.


\end{document}